# Anisotropic Magnetoresistance and Magnetic Anisotropy in High-quality (Ga,Mn)As Films

K. Y. Wang, K. W. Edmonds, R. P. Campion, L. X. Zhao, C.T. Foxon, B.L. Gallagher

*School of Physics and Astronomy, University of Nottingham, Nottingham NG7 2RD, UK*

## Abstract

We have performed a systematic investigation of magnetotransport of a series of as-grown and annealed $Ga_{1-x}Mn_xAs$ samples with $0.011 \leq x \leq 0.09$. We find that the anisotropic magnetoresistance (AMR) generally decreases with increasing magnetic anisotropy, with increasing Mn concentration and on low temperature annealing. We show that the uniaxial magnetic anisotropy can be clearly observed from AMR for the samples with $x \geq 0.02$. This becomes the dominant anisotropy at elevated temperatures, and is shown to rotate by $90^o$ on annealing. We find that the in-plane longitudinal resistivity depends not only on the relative angle between magnetization and current direction, but also on the relative angle between magnetization and the main crystalline axes. The latter term becomes much smaller after low temperature annealing. The planar Hall effect is in good agreement with the measured AMR indicating the sample is approximately in a single domain state throughout most of the magnetisation reversal, with a two-step magnetisation jump ascribed to domain wall nucleation and propagation.

## Introduction

The development of III-V magnetic semiconductors with ferromagnetic transition temperature $T_C$ well in excess of 100K has prompted much interest. The most widely studied material in this category is $Ga_{1-x}Mn_xAs$, with $x$~0.01-0.1, where the randomly-distributed substitutional Mn impurities are ferromagnetically ordered due to interactions with polarised itinerant valence band electrons (holes). The hole density influences all of the magnetic properties of this system, including $T_C$ [1], the magnetic anisotropy [2,3], and the magneto-optical response [4]. There is consequently a strong interplay between magnetic and transport properties [5].

The Giant Magnetoresistance effect and related phenomena in magnetic metal films have found widespread applications in magnetic sensing and recording technologies. Magnetoresistive devices based on III-V magnetic semiconductors may offer a number of advantages over their metallic counterparts: the spin polarisation may be very high [6], suggesting the possibility of larger magnetoresistance effects; the low concentration of magnetic impurities means that fringing fields are weak; magnetic properties may be controllable by dynamic manipulation of the charge carriers [7]; and the technologies for producing III-V semiconductor heterostructures with atomically precise interfaces are well established. Already, a 290% GMR effect in vertical transport [8], and a 2000% in-plane magnetoresistance [9], have been demonstrated in GaMnAs-based devices.

In order to understand and optimise the magnetoresistance of such heterostructures and nanostructures, it is important to develop an improved understanding of the magnetotransport and magnetic anisotropy of single GaMnAs layers. Anisotropic magnetoresistance (AMR) and related effects have been observed in GaMnAs [10,11,12], which are large enough to obscure effects related to spin injection or accumulation in devices. GaMnAs films also show a remarkable variety of magnetic anisotropies. In general, compressive and tensile strained films show in-plane and perpendicular anisotropies respectively, although this also can depend on the hole density. The AMR and the magnetic anisotropy in magnetic materials are intrinsically related to the spin-orbit interaction. In GaMnAs, the substitutional Mn is in a $d^5$ high-spin state, with zero orbital moment. The anisotropy effects are therefore due to the *p*-*d* interactions between Mn and charge carriers, which reside in the valence band of the host semiconductor, where spin-orbit effects are large.

A detailed study of these effects is therefore a key to understanding the nature of the material. Here we investigate the magnetotransport in a series of as-grown and post-growth annealed GaMnAs films on GaAs(001), with a range of different Mn concentrations.

**Experimental details**

The $Ga_{1-x}Mn_xAs$ films were grown on semi-insulating GaAs(001) substrates by low temperature (180ºC-300ºC) molecular beam epitaxy using $As_2$. For all samples studied, the layer structure is 50nm $Ga_{1-x}Mn_xAs$ / 50nm LT-GaAs / 100nm GaAs / GaAs(001). The growth temperature of the $Ga_{1-x}Mn_xAs$ film and the LT-GaAs buffer was decreased with increasing Mn concentration, in order to maintain 2D growth as monitored by RHEED [13]. The Mn concentration was determined from the Mn/Ga flux ratio, calibrated by secondary ion mass spectrometry (SIMS) measurements on 1μm thick films, and includes both substitutional and interstitial Mn. Some of the samples were annealed in air at 190ºC for 50-150 hours, while monitoring the electrical resistance [14]. This procedure has been shown to lead to a surface segregation of compensating interstitial Mn [15,16], and thus can give marked increase of the hole concentration *p* and Curie temperature $T_C$ [17]. X-ray diffraction measurements show that the 50nm films are fully compressively strained, with a relaxed lattice constant *a* that varies linearly with the Mn concentration, as $a=5.65368(1-x)+5.98x$ in the as-grown films, and $a=5.65368(1-x)+5.87x$ after annealing [18]. Full details of the growth and structural characterisation [13], as well as *p* and $T_C$ as a function of Mn concentration [19] are presented elsewhere .

The samples were made into photolithographically defined Hall bars, of width 200μm, with voltage probes separated by 400 μm, and with the current direction along one of the <110> directions. The insulating $x$=0.011 sample discussed below was measured in a van der Pauw geometry, since the very high series resistance of the Hall bar at low temperatures did not permit accurate measurements. In some cases, L-shape Hall bars were used, in which it is possible to measure the magnetoresistance for the current along either the [110] or the $[1\bar{1}0]$ directions. The longitudinal resistance $R_{xx}$ and Hall resistance $R_{xy}$ were measured simultaneously using low frequency ac lock-in techniques. In discussing the results for both types of Hall bars, we define the current direction as $x$, the direction in-plane and perpendicular to the current as $y$, and the growth direction as $z$.

## Results & Discussion

### *I. Anisotropic magnetoresistance in as-grown and annealed GaMnAs*

GaMnAs films are known to show an insulator-to-metal transition with increasing Mn, occurring at around $x$=0.03 in the earliest reports [20], and at lower concentrations in more recent studies [21]. Ferromagnetism can be observed on either side of the transition [20]. In the samples discussed here, the $x$=0.011 film is on the insulating side of the transition, while the other samples studied all show metallic behaviour.

The magnetic field dependence of the sheet resistance at sample temperature T=4.2 K, for a series of as-grown and annealed $Ga_{1-x}Mn_xAs$ thin films with $x$ between 0.011 and 0.067, are shown in Fig.1. For all samples, two contributions to the magnetoresistance can be distinguished. At fields greater than the saturation magnetic field, a negative magnetoresistance is observed, the slope of which is independent of the external field direction. This isotropic magnetoresistance does not saturate even for applied fields above 20T [22], and has been attributed to suppression of weak localisation and spin-disorder scattering at low and high temperatures respectively [22,23,24]. The isotropic magnetoresistance becomes weaker after low temperature annealing after removing the compensating defects. The second contribution occurs at lower fields, and is dependent on the field orientation. This is the anisotropic magnetoresistance which is the subject of this paper. As a result of the spin-orbit interaction and its effect on scattering between carriers and magnetic ions, the resistivity depends on the angle between the sample magnetisation and the applied current. This is a well-known effect in ferromagnetic materials. Applying a small magnetic field leads to rotation of the magnetisation into the field direction, which gives rise to the low-field magnetoresistance effects shown in fig. 1.

The low-field magnetoresistance traces are qualitatively similar to those reported elsewhere for GaMnAs thin films [10,11], and yield information concerning the magnetic anisotropy. For all samples, the resistance at zero field is independent of the angle of the previously applied field, indicating that the magnetisation always returns to the easy axis on reducing the field to zero. For most of the films, the lowest resistance state is obtained when H is along the x-direction, while the field where the AMR saturates is largest for H along the z-direction, indicating that this is a hard magnetic axis.

Significantly different behaviour can be observed between the sample with $x$=0.011 and the other samples, i.e. between samples lying on either side of the metal-insulator transition. For $x$=0.011, the resistance is largest for in-plane magnetic field. This is usually the case for ferromagnetic metals, but is opposite to what is observed for the metallic GaMnAs films. In addition, the saturation field obtained from the AMR is larger for fields applied in-plane than for fields out-of-plane, which indicates that this sample possesses a perpendicular magnetic anisotropy. It has been noted previously that for compressive-strained GaMnAs films at low hole concentrations the easy magnetic axis can lie perpendicular to the plane [25]. The present result shows that both the magnetic anisotropy and the anisotropic magnetoresistance are of opposite sign in the $x$=0.011 sample, as compared to the metallic samples. The sample with $x$=0.017 appears to be an intermediate case, where the low resistance state is for in-plane magnetisation, while in-plane and out-of-plane saturation fields are of comparable magnitude.

The saturation field for H||z and H||y for the as-grown and annealed samples with $x \geq 0.017$ is shown in fig.2 (a) and (b), respectively. With increasing Mn concentration, the saturation field for in-plane (out-of-plane) directions becomes smaller (larger) for the as-grown samples, i.e. the in-plane magnetic anisotropy becomes weaker. On annealing, the in-plane saturation field does not change in a

systematic way or vary monotonically with Mn concentration. The easy magnetic axis is defined by a competition between the uniaxial anisotropy between [110] and [$1\bar{1}0$] directions, $K_u$, and a biaxial anisotropy $K_b$ which favours orientation of the magnetisation along the in-plane <100> directions. At low temperatures with $K_b > K_u$, the easy axis will lie in the direction $\frac{a\cos(-K_u / K_b)}{2}$ away the uniaxial easy axis towards the cubic easy axis [28]. The saturation magnetic field along y direction is dependent on competition of these two magnetic anisotropies, while the saturation magnetic field for H out-of-plane becomes significantly larger, i.e. the z-axis becomes significantly harder. The principal effect of annealing is to increase the hole density, through out-diffusion of compensating Mn interstitial defects [15,16]. The magnetic anisotropy in III-V magnetic semiconductors is well explained within the Zener mean field model, which predicts that the in-plane anisotropy field increases with increasing hole density and compressive strain [2]. The trends observed on increasing the Mn concentration and on annealing are in agreement with this prediction.

Since both the AMR and the magnetic anisotropy originate from the spin-orbit interaction, a close correlation between the two effects may be expected, as is demonstrated here. We quantify the AMR for magnetisation parallel and perpendicular to the plane as respectively,

$AMR_{//} =(R_{//x}-R_{//y})*100/R_{//x}(\%)$ and

$AMR_{\perp} = (R_{//x}-R_{//z})*100/R_{//x}(\%)$,

with $R_{//i}$ the sheet resistances for magnetisation parallel to the $i(=x,y,z)$ axis. These are plotted in fig. 3 (a) and (b) for samples with $0.017 \leq x \leq 0.09$ before and after annealing, at temperature 4.2K and at the saturation field. For the as-grown samples, both $AMR_{//}$ and $AMR_{\perp}$ generally decrease with increasing Mn, while the difference between $AMR_{//}$ and $AMR_{\perp}$ generally increases. The AMR decreases slightly after annealing, even though the resistivity has decreased, i.e. the absolute value of ΔR decreases significantly. The data of fig. 3(a) has been quantitatively described within a model of band-hole quasiparticles with a finite spectral width due to elastic scattering from Mn and compensating defects, using known values for the hole density and compressive strain, and no free parameters, presented elsewhere [5]. From fig 3(a) and (b), it can be seen that the AMR generally decreases while the magnetic anisotropy increases, both with increasing Mn and on annealing. A similar trend of increasing AMR with decreasing magnetic anisotropy is observed in metallic magnetic compounds, e.g. the NiFe system[26].

The ratio $AMR_{\perp}/AMR_{//}$ is plotted in fig. 3(c), and very different behaviour is observed for samples before and after annealing. Before annealing, $AMR_{\perp}$ is up to a factor of two larger than $AMR_{//}$, and the ratio systematically increases with increasing Mn concentration. After annealing, the ratio is comparable to or less than 1 for all concentrations. The origin of this difference between in-plane and out-of-plane AMR is not clear, however the precise nature of the AMR and magnetic anisotropy is likely to depend on a detailed balance between strain and the concentration of holes, Mn, and other defects, all of which may be affected by annealing.

The effect of annealing on the AMR, the ratio $AMR_{\perp}/AMR_{//}$, and the saturation field becomes progressively less pronounced with decreasing $x$, until at $x$=0.017 where almost no change is observed. A decreasing effect of annealing with decreasing $x$ is also observed for the hole density as well as $T_C$, which indicates that the number of interstitial Mn is small at low $x$ [19]. With increasing Mn concentration, there is an increasing tendency for the Mn to auto-compensate by occupying interstitial sites.

*II. Uniaxial magnetic anisotropy*

For the annealed sample with *x*=0.067, the sheet resistance sharply increases on applying a small magnetic field in y direction, while no magnetoresistance is observed for H applied along the *x* direction, as shown in figure 1h. This indicates that the magnetic moment is oriented either parallel or antiparallel to this direction throughout the whole magnetisation reversal, in turn indicating the presence of a dominant in-plane uniaxial magnetic anisotropy. A uniaxial magnetic anisotropy between the in-plane [110] and $[1\bar{1}0]$ directions in GaMnAs has been noted previously [10,12,27,28], and is observed to some degree in all the samples discussed in the present study.

In compressive strained GaMnAs films, magnetic domains can be very large, extending over several mm [28], and at remanence the films tend to lie in a single-domain state [29]. If $K_u>K_b$, then the magnetisation at H=0 is fixed along the easier of the <110> directions, whereas if $K_u<K_b$, the magnetisation at H=0 is oriented between the <100> and <110> directions, moving closer to <100> as $K_b$ becomes larger. The former appears to be the case for the annealed *x*=0.067 sample. For the other metallic samples shown in figures 1, the resistance at H=0 is intermediate between its saturation values for H//x and H//y, indicating that $K_b>K_u$ for these samples at T=4.2K. Since $K_b$ and $K_u$ are proportional to $M^4$ and $M^2$ respectively, where M is the magnetisation, the former falls more rapidly with increasing temperature than the latter. Therefore, with increasing temperature, the easy magnetic axis rotates away from the <100> directions. This has been observed directly using magneto-optical imaging [28], and can also be inferred from analysis of the temperature-dependence of the remnant magnetisation measured by SQUID [29]. This rotation can also be seen in the AMR. Figure 4 a and b show the AMR for the as-grown $x$ = 0.034 sample measured for different in-plane field orientations at T = 4.2K and T = 40K, respectively. At both temperatures, the low-field magnetoresistance is largest for H//x. The other two orientations show similar magnetoresistance at 4.2K, No magnetoresistance (aside from the isotropic negative slope seen for all orientations) is observed for H//y at 40 K. The angle-dependent diagonal component of the resistivity tensor under a single domain model is given by:

$$\rho_{xx}(\theta) = \rho_{//}\cos^2\theta + \rho_{\perp}\sin^2\theta = (\rho_{//}+\rho_{\perp})/2 + \tfrac{1}{2}(\rho_{//}-\rho_{\perp})\cos 2\theta = \rho_0 + \Delta\rho\cos 2\theta \qquad (1)$$

where θ is the angle between magnetisation and current direction (along [110] direction for this sample ). Rearranging Equation (1), we can get:

$$\theta = \frac{1}{2}a\cos\left(\frac{\rho_{//} + \rho_{\perp} - 2\rho_{xx}(\theta)}{\rho_{\perp} - \rho_{//}}\right) \qquad (2)$$

Inserting the zero magnetic field resistivity as $\rho_{xx}(\theta)$ of Equation (2), the magnetization direction is obtained. The easy axis at 4.2 K is between [100] and $[1\bar{1}0]$ directions and is 22±4$^0$ away from $[1\bar{1}0]$ direction, which is consistent with our magnetometry results. With increasing temperature, the uniaxial magnetic anisotropy is dominant, and the magnetisation is locked parallel or antiparallel to the y direction, consistent with the magnetometry studies [29].

By comparing SQUID magnetometry results with Laue back-reflection and RHEED measurements, we have shown elsewhere that the uniaxial easy axis is along the $[1\bar{1}0]$ direction in all the as-grown samples studied by us [30]. On annealing samples with $x \geq 0.04$, the easy axis is found to rotate by 90° into the [110] direction. This can also be observed in the AMR response, by comparing figures 1e and h, which correspond to the same $x$=0.067 Hall bar before and after annealing. Figure 1h shows that the easy axis is aligned along the $x$-direction for this sample after annealing. Before annealing, a low-field magnetoresistance is observed both for B//x and H//y, indicating that the easy axis is close to 45° from the <110> axes at this temperature, and the biaxial anisotropy is dominant. However, it can be seen that the largest magnetoresistance is observed for H//x, which means that the easy axis is slightly tilted towards the direction perpendicular to the current. Therefore, in the as-grown film the *y*-direction is the easier of the two <110> axes. Etching studies show that this 90º rotation of the uniaxial easy axis is not related to Mn surface-segregation [30], and is likely to be due to the increased hole density and the influence of this on the magnetic anisotropy.

To further investigate the uniaxial magnetic anisotropy and its influence on the AMR, we also performed measurements on L-shaped Hall bars, in which the current is parallel to the [110] direction along one branch, and parallel to the $[1\bar{1}0]$ direction along the other. The magnetoresistance for current along the two arms, for $x = 0.034$ and T=4.2K, is shown in figure 5a and 5b. Along arm 'a', the resistivity is initially relatively low, and increases to a high value when a magnetic field is applied perpendicular to the current direction, either in- or out-of-plane. In contrast, along arm 'b', the resistance change is largest when the field is applied parallel or antiparallel to the current. This demonstrates that the easy magnetic axis lies close to the same <110> direction in both arms of the Hall bar. It is also worth noting that both $AMR_{//}$ and $AMR_{\perp}$ are around 20% larger along arm 'b' than along arm 'a'. This may reflect a dependence of the AMR on the angle between the current / magnetisation and certain crystallographic axes, as well as their relative orientation, as will be discussed in the next section.

*III. Planar Hall effect*

The combination of an AMR effect of several percent and a large absolute value of the sheet resistance gives rise to a giant 'planar Hall effect' in GaMnAs, which has been studied in detail elsewhere [10]. This effect arises as a result of the non-equivalence of components of the resistivity tensor which are perpendicular and parallel to the magnetisation direction, leading to the appearance of off-diagonal resistivity components. The angle-dependent off-diagonal component of the resistivity tensor under a single domain model are given by:

$$\rho_{xy}(\theta) = (\rho_{//}-\rho_{\perp})\cos\theta\sin\theta = ½(\rho_{//}-\rho_{\perp})\sin2\theta = \Delta\rho\sin2\theta \quad (3)$$

where θ is the angle between magnetisation and current. In fig. 6 (a) and (b), we show longitudinal and planar Hall resistivities for the as-grown $x$=0.034 sample, measured while rotating a 0.6T external magnetic field in the plane of the Hall bar. As expected from the above relationships, the planar Hall resistivity is largest when the field is at 45° to the current direction, and zero for field and current parallel or perpendicular. However, fitting the data of figure 6 to equations (1) and (3) yields only qualitative agreement. The amplitude of the Hall oscillation is found to be smaller than the value of Δρ obtained from the longitudinal resistivity measurements. Also, the shape of the longitudinal resistivity oscillation shows some deviations from a cos2θ dependence on field angle. We obtain a much better fit by adding an additional term $\rho_1\cos4\theta$ to equation (1). The best fit to the angle dependent resistivity yields, Δρ= -90μΩcm, and $\rho_1$= -12 μΩcm. The $\rho_1$ term reflects a magnetocrystalline contribution to the resistivity when the magnetization is directed away from the main crystalline axes. A similar 4$^{th}$ order term was recently identified in the AMR response of epitaxial Fe(110) films [31]. This 4$^{th}$ order term is not observed in the Hall resistivity because the magnetocrystalline contribution to the Hall resistivity under cubic symmetry is 2nd order [32]. The 4th order term in $\rho_{xx}$ is typically around 10-15% of the 2nd order term in the as-grown films. After annealing, the 4th order term becomes much smaller, and the angle-dependent resistivities can be described approximately by equations (1) and (3). However, we find that the amplitude of the oscillations of $\rho_{xx}$ is larger than that of $\rho_{xy}$ by a factor of 1.3 for this sample. This value is sample-dependent may be due to a difference in the AMR in the Hall cross region compared to the region between the crosses.

Figure 7 shows the anisotropic magnetoresistance and planar Hall effect versus external magnetic field, applied along various in-plane directions, for the as-grown $x$=0.034 film at 4.2K. At $\theta$=±45°, the planar Hall trace is qualitatively similar to those presented in ref. [12], showing sharp hysteretic spikes at around 25mT. More complicated behaviour is observed when the magnetic field is applied parallel or perpendicular to the current direction. For these orientations, the spikes are much broader, and are superimposed on a slowly varying background. The anisotropy between the in-plane <110> directions can be clearly seen by comparing the width of both the spikes and the background feature for the two orientations.

Equations (1) and (3) can be rearranged to give

$$\rho_{xy}(\theta) = \tfrac{1}{2}(\rho_{//} - \rho_{\perp})\left[1 - \left(\frac{2\rho_{xx}(\theta) - \rho_{//} - \rho_{\perp}}{\rho_{//} - \rho_{\perp}}\right)^2\right]^{1/2} \qquad (4)$$

The square root can take positive or negative values, depending on the magnetisation angle $\theta$. Inserting the measured values of $\rho_{\mathrm{xx}}$ into the equation (4) allows us to predict the value of $\rho_{\mathrm{xy}}$ for a given external magnetic field. The measured and predicted field dependence of $\rho_{\mathrm{xy}}$ are shown figure 7(b-e). Here we have reduced the measured $\rho_{xx}$ by the factor of 1.3 to allow for the experimental difference in overall magnitude discussed above. The predicted results are in good agreement with the measurement except for the larger values of $90^0$ case, provided that the sign of the square root in equation (4) is chosen correctly. This indicates the sample remains approximately in a single domain state throughout the magnetisation reversal.

Since $\rho_{xy}$ can be described according equation (4), this can also be used to determine the field dependence of the magnetisation angle θ. This is shown in figure 8a and b, for external magnetic field along $\theta = 0^0$ and $45^0$ respectively. For both orientations, θ shows sharp jumps at two distinct fields for each sweep direction, together with regions where θ is slowly varying. The jumps are large and closely spaced in H for $\theta=45^0$, and smaller and more widely spaced for $\theta=0^0$. The jumps are ascribed to nucleation and propagation of domain walls which occur over a narrow field range, as is observed elsewhere [28]. Away from the jumps, the planar Hall resistivity is well-described by equations (1) and (3) (figure 7), indicating that the sample is approximately in a single-domain state, and the slow variation of θ is ascribed to coherent rotation. The magnetisation does not directly reverse even for $\theta = 45^0$, which is a consequence of the coexisting biaxial and uniaxial in-plane magnetic anisotropies [12].

**Summary**

The AMR for a series of as-grown and annealed (Ga,Mn)As samples has been carefully studied. Both $AMR_{//}$ and $AMR_{\perp}$ generally decrease with increasing Mn for the as-grown samples. $AMR_{\perp}$ is up to a factor of two larger than $AMR_{//}$, and the ratio systematically increases with increasing Mn concentration. After annealing, the AMR decreases slightly, the ratio of $AMR_{\perp}/AMR_{//}$ is closer to 1 and decreases slightly with increasing $x$ up to 0.067. The uniaxial magnetic anisotropy could be clearly observed from AMR for the samples with $x \geq 0.022$. The in-plane longitudinal resistivity has contributions not only from the relative angle between magnetization and current direction, but also from the relative angle between magnetization and the main crystalline axes. The latter term becomes much smaller after low temperature annealing. The predicted values of $\rho_{xy}$ are in good agreement with the measurements indicating the sample remains approximately in a single domain state throughout the magnetisation reversal. The predicted values of θ show that the magnetic switching can be understood according to a two step jump by nucleation and propagation of $90^0$ domain walls.

**Acknowledgements**

We are grateful for financial support from the EPSRC (GR/S81407/01). We also thank Jaz Chauhan and Dave Taylor for processing of the Hall bars.

---

## Figure Captions:

Fig.1 Sheet resistance as a function of magnetic field at T = 4.2K for the as-grown $Ga_{1-x}Mn_xAs$ thin films with different value of $x$ (a) $x$ = 0.011, (b) $x$ = 0.017, (c) $x$ = 0.022, (d) $x$ = 0.034 (e) $x$ = 0.067, for the annealed samples with (f) $x$ = 0.022, (g) $x$ = 0.034 and (h) $x$ =0.067, with three mutually orthogonal orientations ( [110], $[1\bar{1}0]$ and [001] directions) of the magnetic field.

Fig.2 The saturation magnetic field on applying (a) H||z ( [001]direction) and (b) H||y for the as-grown and annealed $Ga_{1-x}Mn_xAs$ samples with $0.017 \leq x \leq 0.09$ at 4.2 K.

Fig.3 The $AMR_{//}$ and $AMR_{\perp}$ for (a) the as-grown and (b)annealed $Ga_{1-x}Mn_xAs$ samples with $0.017 \leq x \leq 0.09$ at 4.2 K;(c) the ratio of $AMR_{\perp}/ AMR_{//}$ versus Mn concentration for the as-grown and annealed samples at 4.2 K.

Fig.4 (Colour Online)The in-plane anisotropic magnetoresistance at (a) 4.2 K and (b) 40K for the as-grown $Ga_{1-x}Mn_xAs$ thin film with $x$ = 0.034 when current lies in [110] direction (thin black lines up sweep, thick gray lines down sweep). The easy axis at 40K is clearly along (H < I = $90^0$) $[1\bar{1}0]$ direction because almost no anisotropic magnetoresistance is observed during magnetic reversal along this direction.

Fig.5 (Colour Online)The sheet resistance a function of magnetic field at T = 4.2K for an L shaped sample of $Ga_{1-x}Mn_xAs$ with $x$ = 0.034. (a) current $[1\bar{1}0]$ along direction with three mutually orthogonal orientations of the magnetic field;(b) current along

[110] direction with three mutually orthogonal orientations of the magnetic field. (in both graphes the thick gray lines up sweep, thin black lines down sweep).

Fig.6 The angular dependence of (a) $\rho$-$\rho_0$ ($\rho_0 =(\rho_{//}+\rho_{\perp})/2$) and (b) Hall resistivity for the as-grown $Ga_{1-x}Mn_xAs$ with $x$ = 0.034 thin film under the external magnetic field H = 6000 Oe at 4.2 K, the solid lines are best fitting results.

Fig.7 (Colour Online) (a)The sheet resistance as a function of in-plane magnetic field at 4.2K with different angles. (b) to (e) Measured (open triangles up sweep, closed circles down sweep) and predicted (thin black lines up sweep, thick gray lines down sweep) Hall resistance as a function of in-plane magnetic field at 4.2K at different angles (b)$\theta$ = -$45^0$ (c) $\theta = 0^0$ (d) $\theta = 45^0$ (e) $\theta = 90^0$ for the as-grown $Ga_{1-x}Mn_xAs$ thin films with $x$ = 0.034.

Fig.8 the predicted magnetization direction $\theta$ vs. external magnetic field when (a)$\theta$ = $0^0$ and(b) $45^0$.

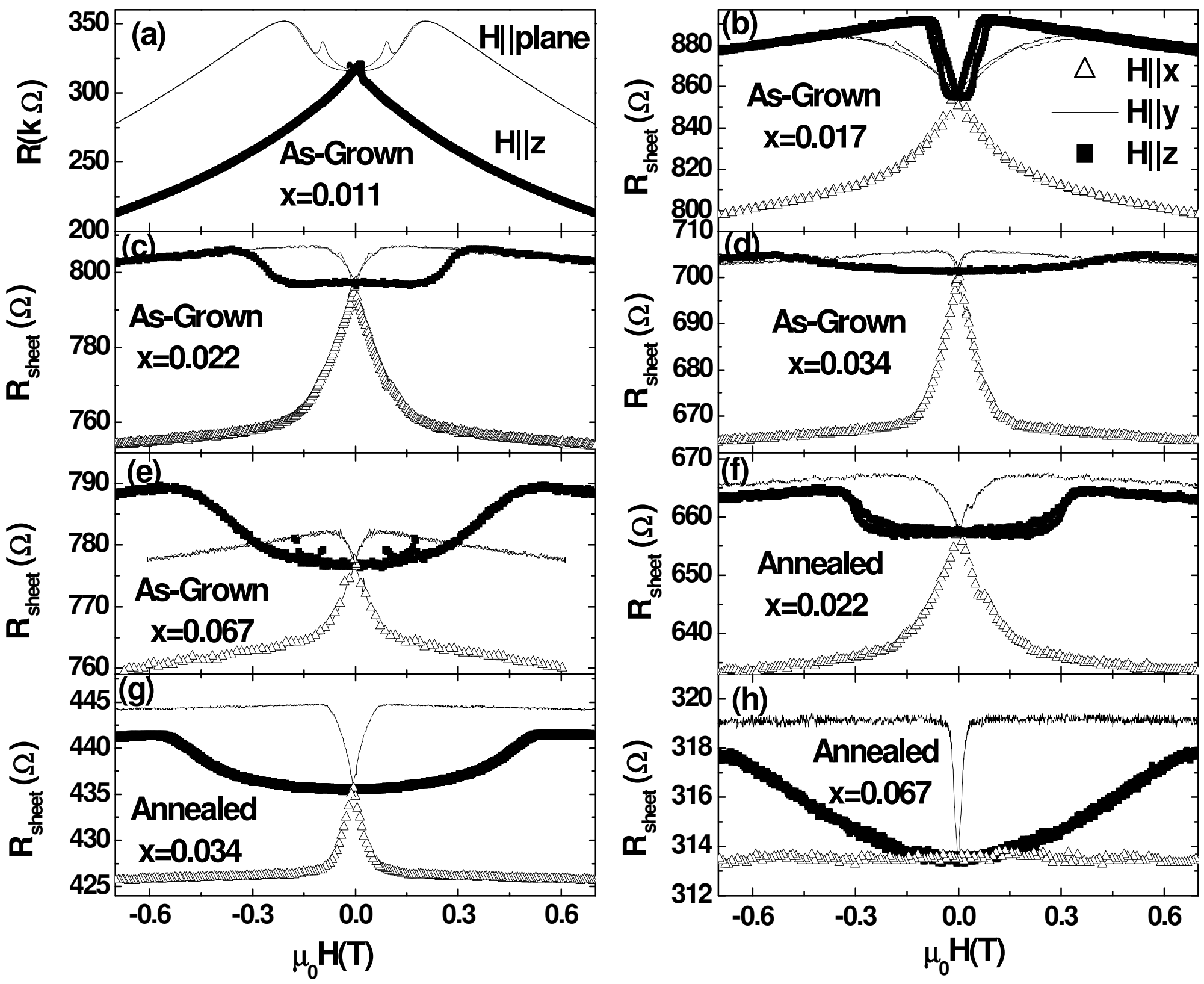

Fig.1 K. Y. Wang et al.

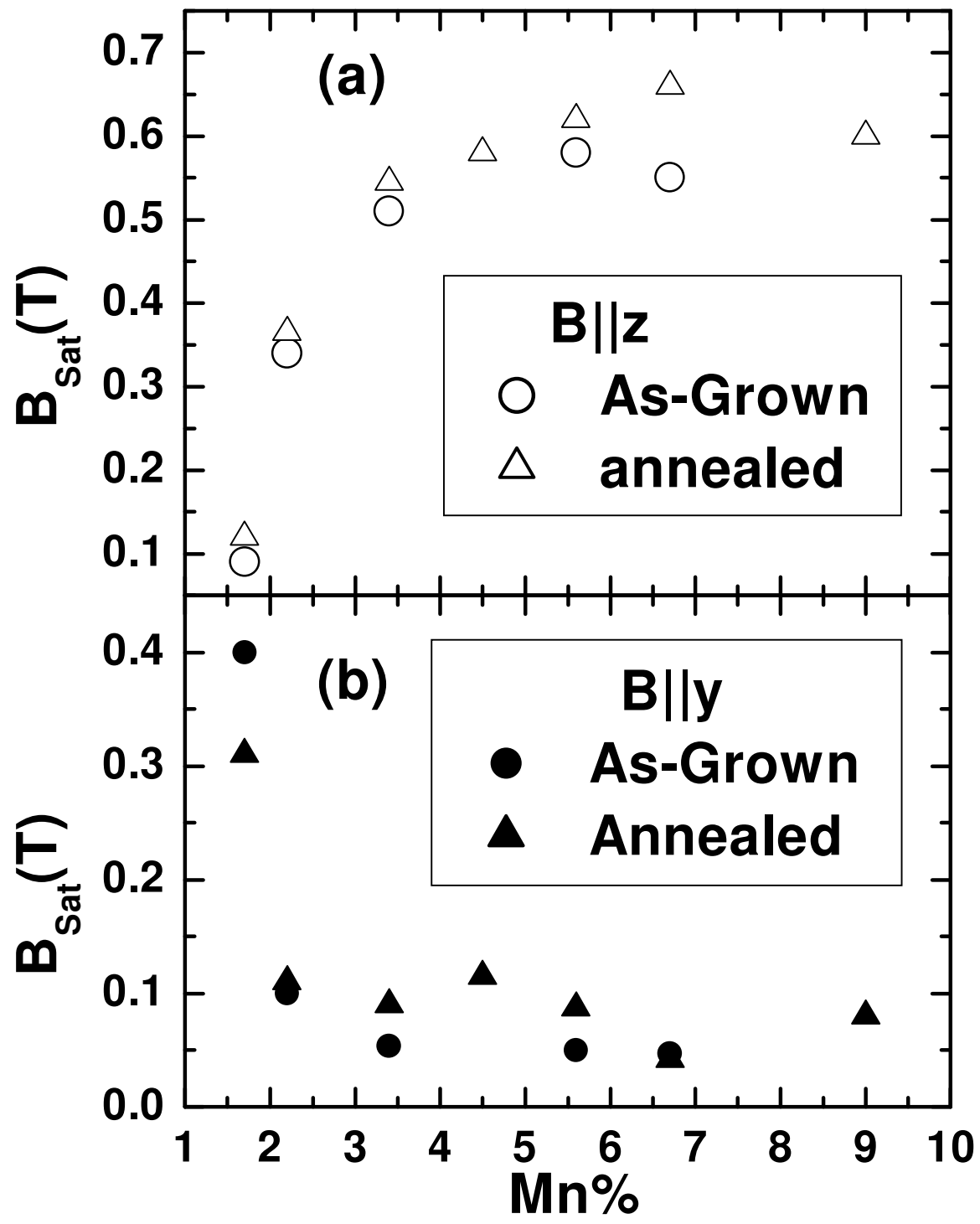

Fig.2 K. Y. Wang et al.

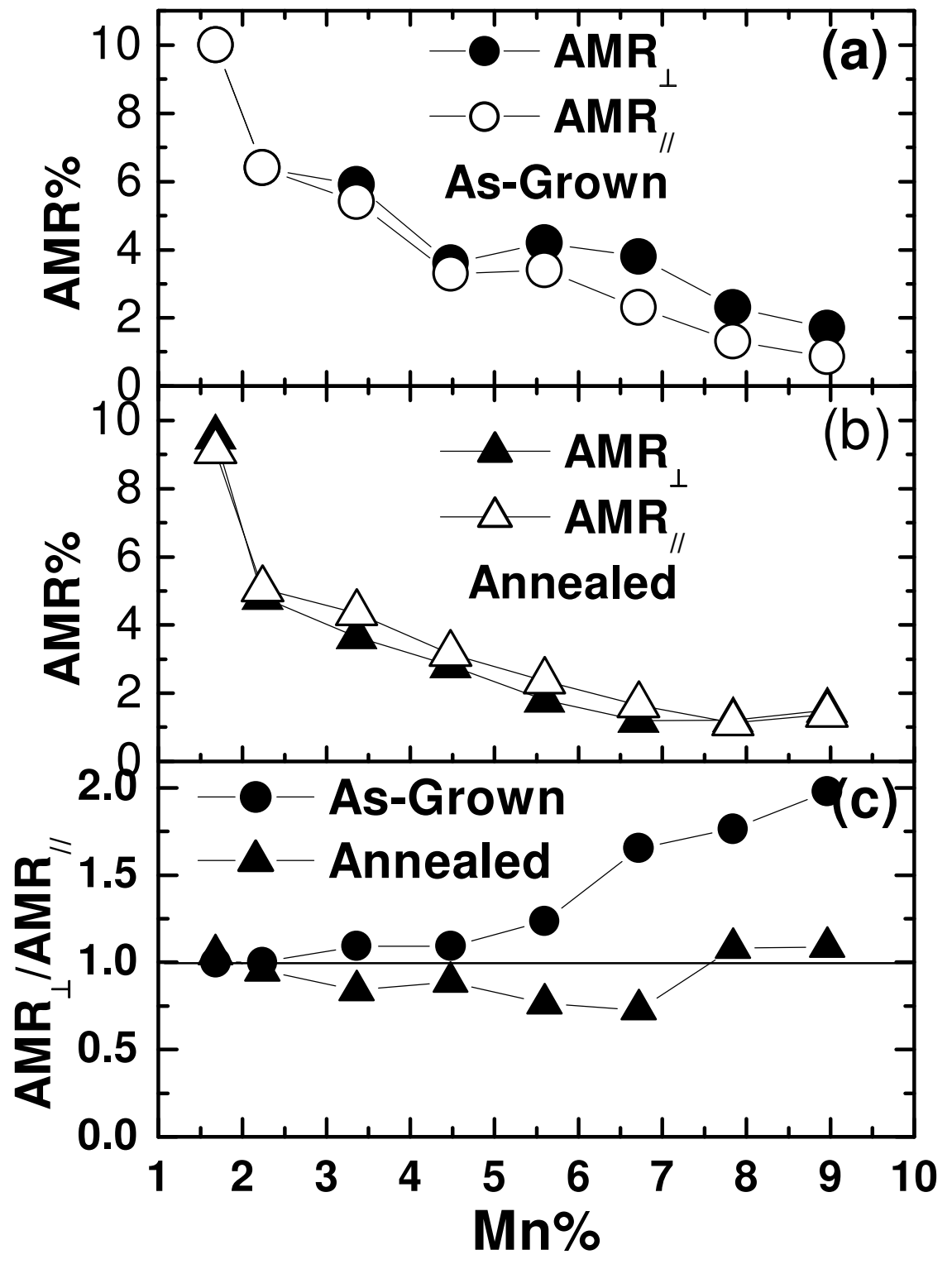

Fig.3 K. Y. Wang et al.

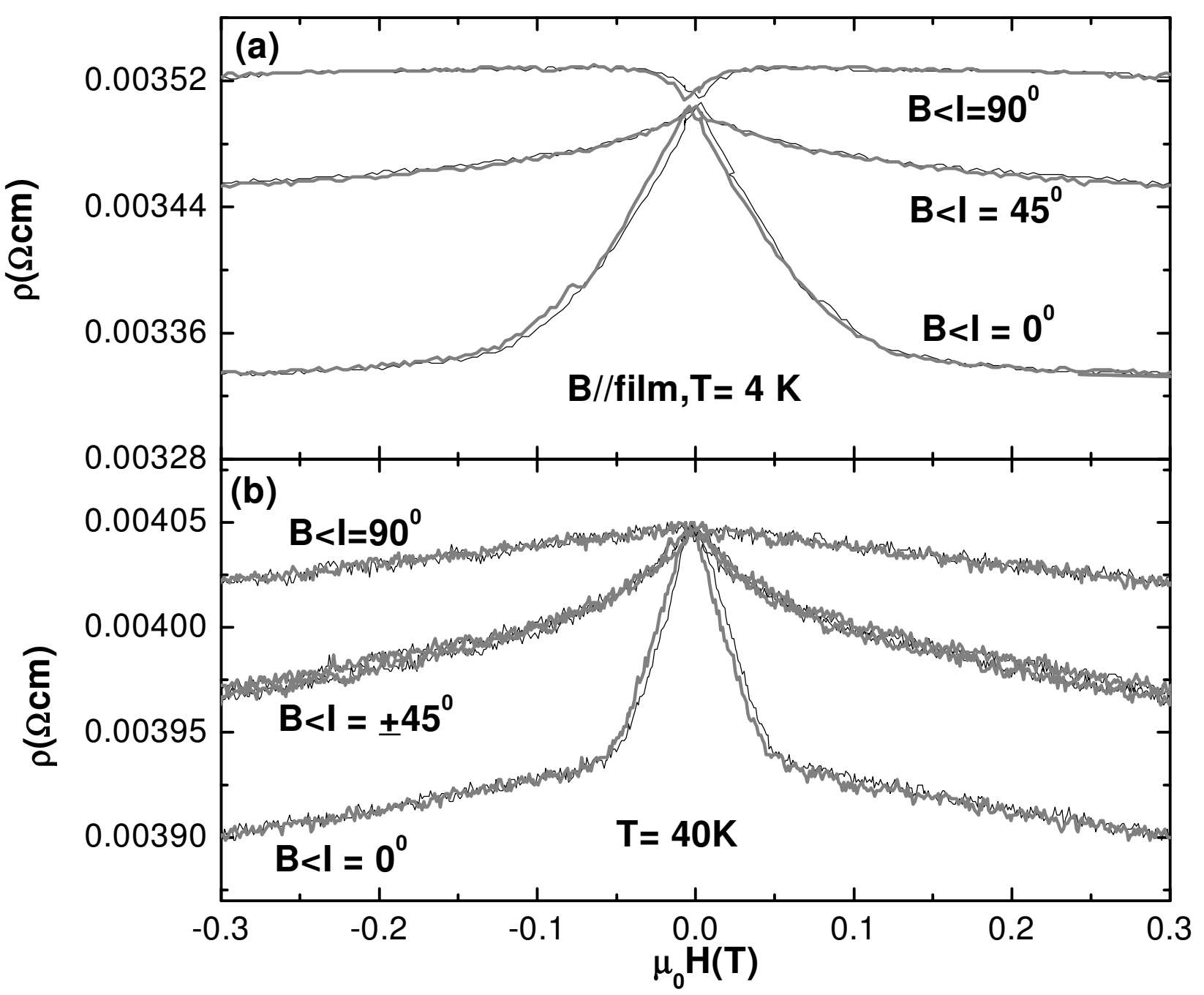

Fig.4 K. Y. Wang et al.

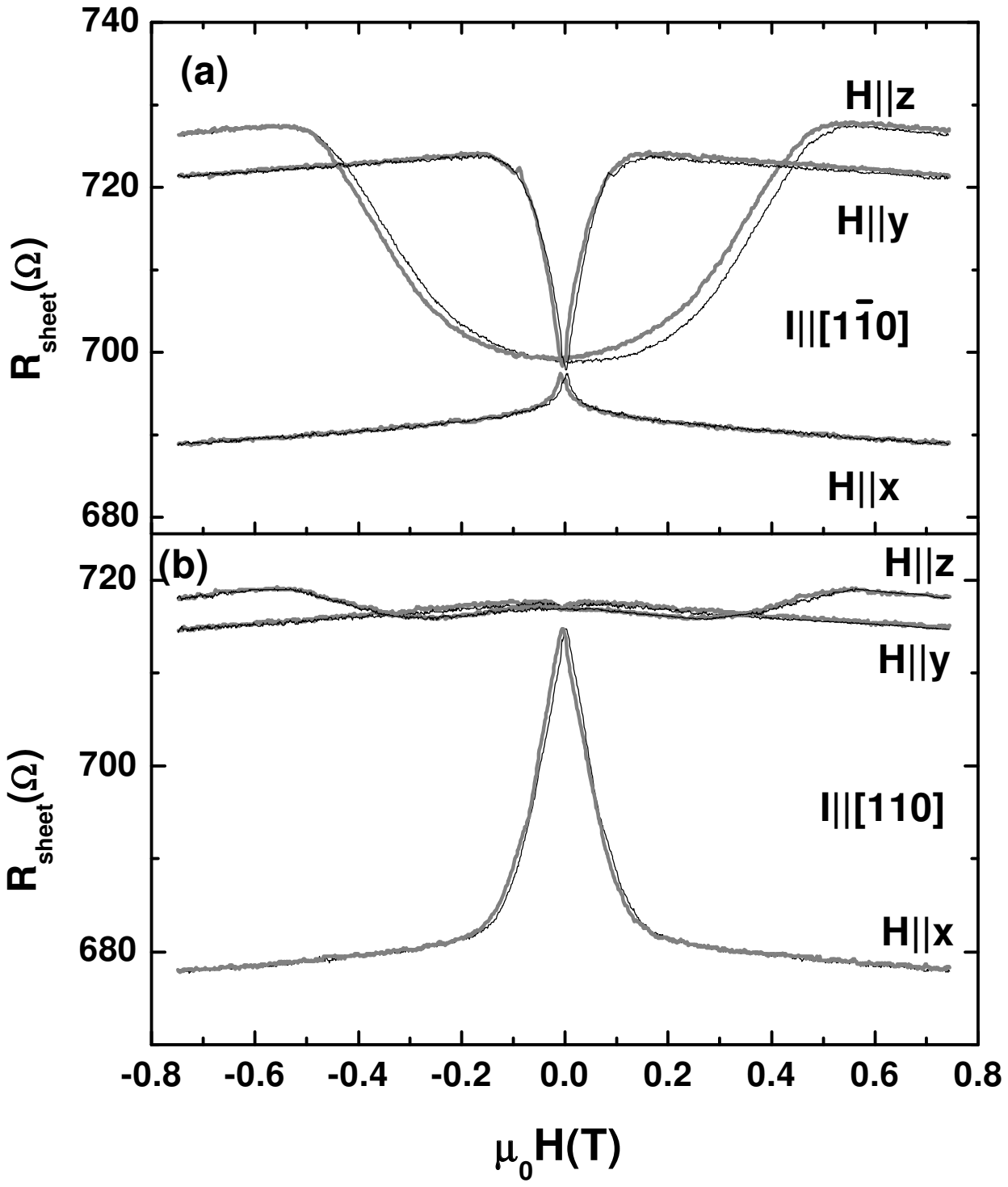

Fig.5 K. Y. Wang et al.

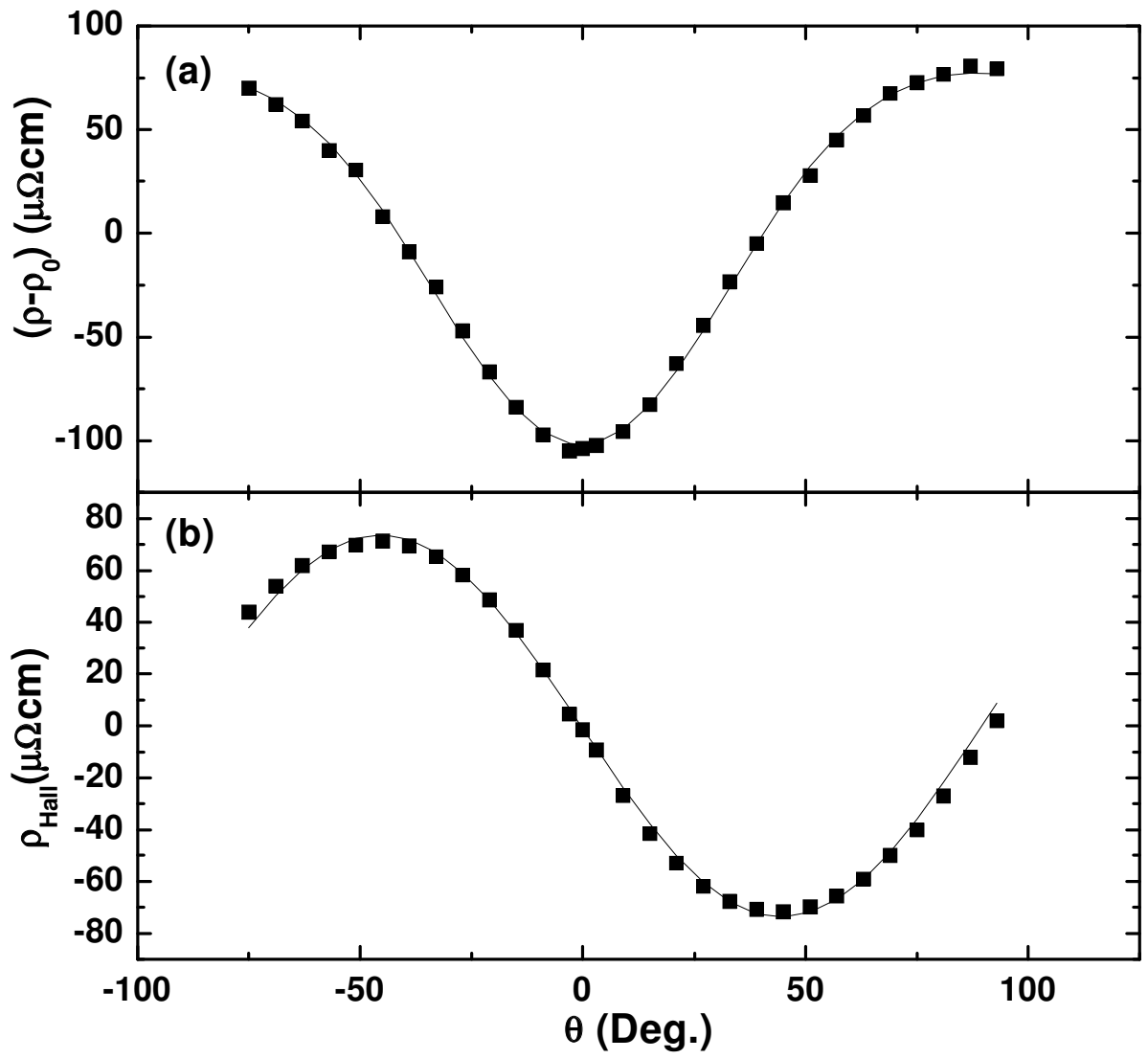

Fig.6 K. Y. Wang et al.

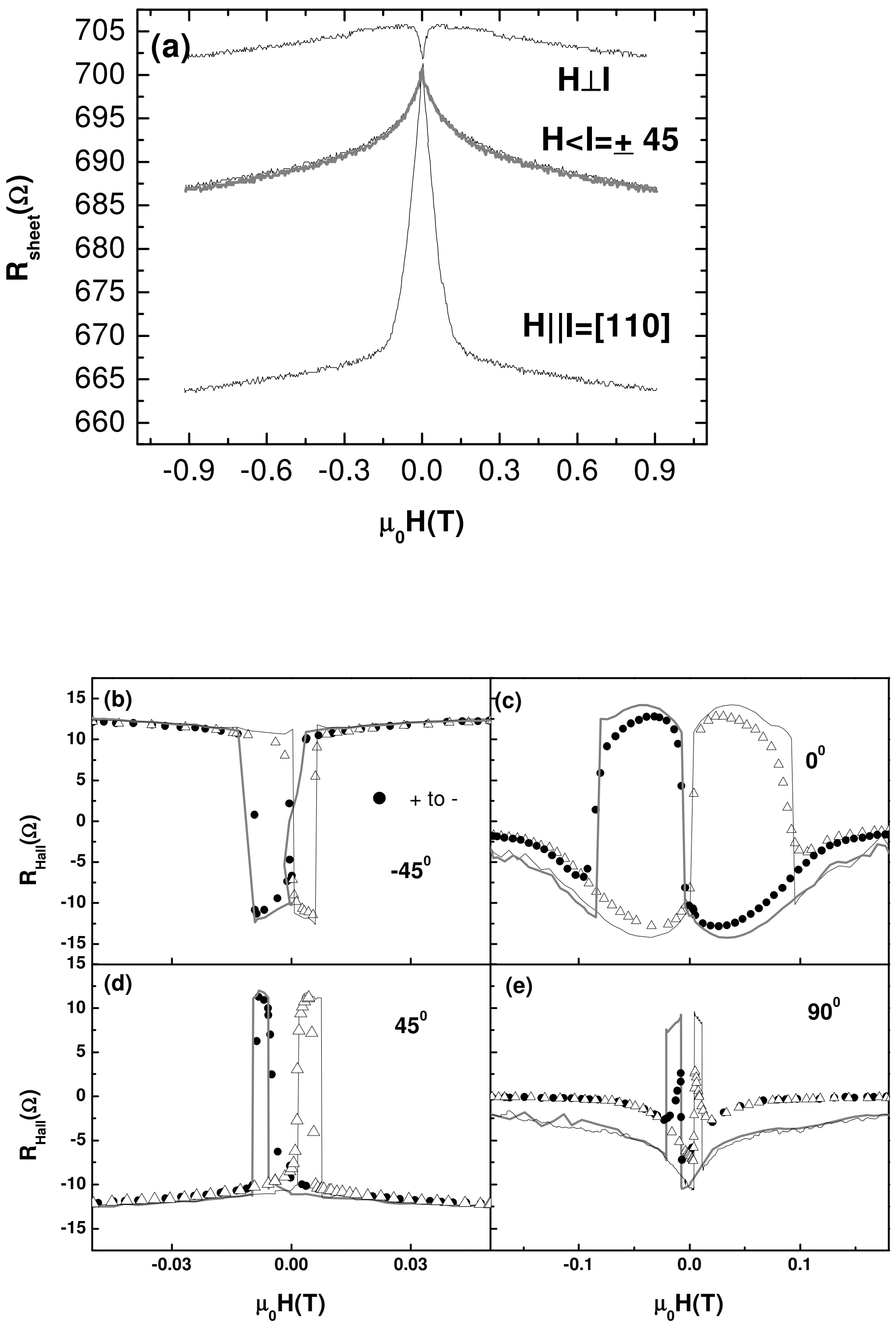

Fig.7 K. Y. Wang et al.

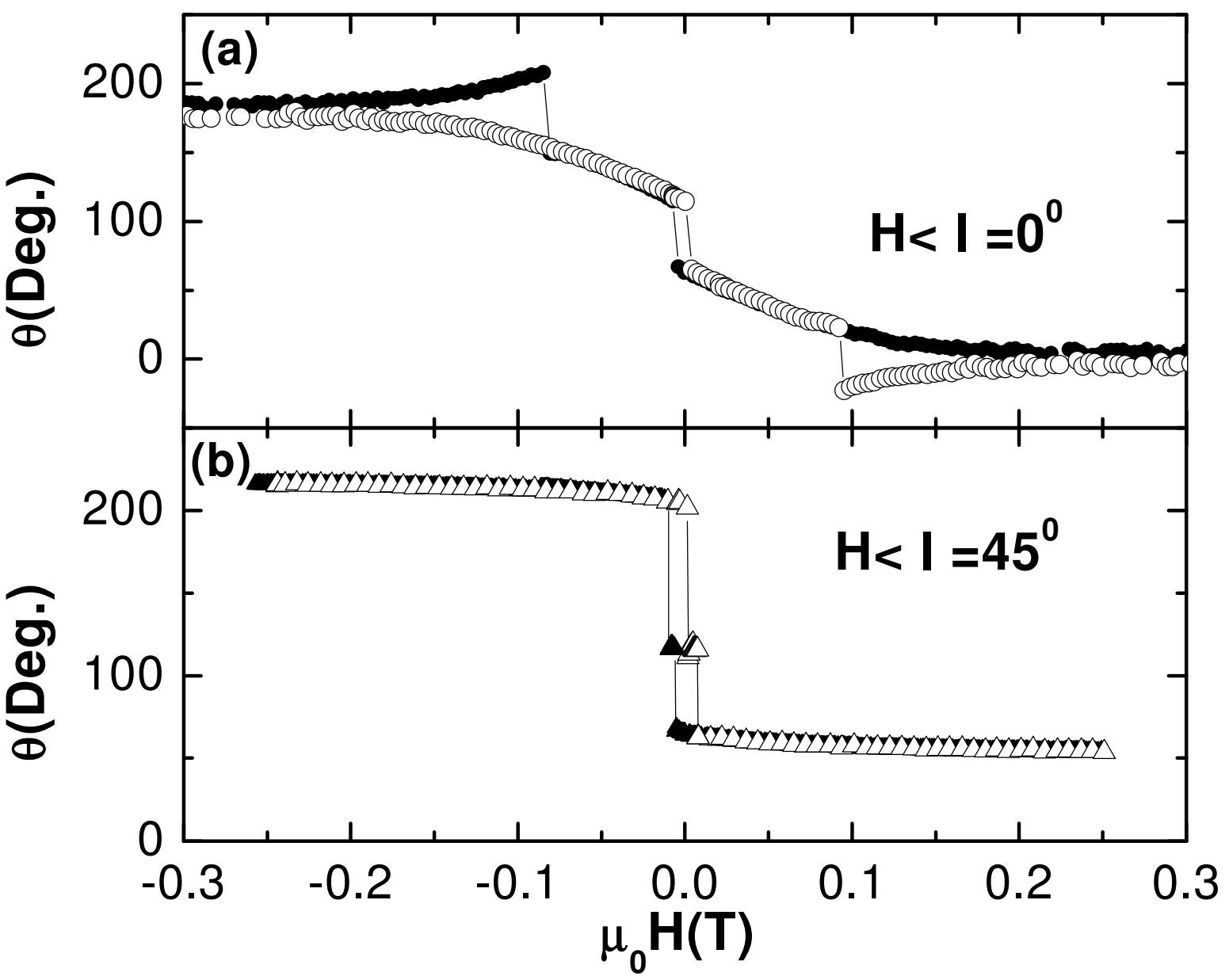

Fig.8 K. Y. Wang et al.